\documentclass[lettersize,journal]{IEEEtran}
\usepackage{amsmath,amsfonts}
\usepackage{algorithmic}
\usepackage{algorithm}
\usepackage{array}
\usepackage[caption=false,font=normalsize,labelfont=sf,textfont=sf]{subfig}
\usepackage{textcomp}
\usepackage{stfloats}
\usepackage{url}
\usepackage{xcolor}
\usepackage{verbatim}
\usepackage{graphicx}
\usepackage{cite}
\usepackage{xcolor}
\usepackage{tabularx,booktabs,textcomp}
\usepackage{amsthm}

\usepackage{amsmath,amssymb,amsthm}

\begin{document}

\title{Evaluation of Energy Resilience and Cost Benefit in Microgrid with Peer-to-Peer Energy Trading}

\author{{Divyanshi Dwivedi, K. Victor Sam Moses Babu,~\IEEEmembership{Member,~IEEE}, Pradeep Kumar Yemula,~\IEEEmembership{Member,~IEEE}, Pratyush Chakraborty,~\IEEEmembership{Member,~IEEE}, Mayukha Pal,~\IEEEmembership{Senior Member,~IEEE}}
 

\thanks{(Corresponding author: Mayukha Pal)}

\thanks{Mrs. Divyanshi Dwivedi is a Data Science Research Intern at ABB Ability Innovation Center, Hyderabad 500084, India and also a Research Scholar at Department of Electrical Engineering, Indian Institute of Technology, Hyderabad 502205, IN, (e-mail: divyanshi.dwivedi@in.abb.com).}
\thanks{Mr. K. Victor Sam Moses Babu is a Data Science Research Intern at ABB Ability Innovation Center, Hyderabad 500084, India and also a Research Scholar at the Department of Electrical and Electronics Engineering, BITS Pilani Hyderabad Campus, Hyderabad 500078, IN, (e-mail: victor.babu@in.abb.com).}
\thanks{Dr. Pradeep Kumar Yemula is an Assoc. Professor with the Department of Electrical Engineering, Indian Institute of Technology, Hyderabad 502205, IN, (e-mail: ypradeep@ee.iith.ac.in).}
\thanks{Dr. Pratyush Chakraborty is an Asst. Professor with the Department of Electrical and Electronics Engineering, BITS Pilani Hyderabad Campus, Hyderabad 500078, IN, (e-mail:pchakraborty@hyderabad.bits-pilani.ac.in).}
\thanks{Dr. Mayukha Pal is a Global R\&D Leader – Data Science at ABB Ability
Innovation Center, Hyderabad-500084, IN, (e-mail: mayukha.pal@in.abb.com).}

}


\maketitle

\begin{abstract}

Integration of distributed energy resources (DERs) at a residential level has given way to the  development of a self-sufficient energy-resilient system and peer-to-peer (P2P) energy trading at a community level. Several researchers have developed models to implement energy trading in the P2P network with DERs that enables prosumers to avail cost benefits. However, the impact of DERs and P2P energy network on the system's resilience is not evaluated with a quantifiable measure. In this work, we use the percolation threshold as a measure to quantify energy resilience, which can be computed using complex network. We consider a standard IEEE-123 node test feeder integrated with renewable energy sources and partitioned into microgrids. The microgrid with the highest percolation threshold value is considered for P2P energy trading analysis. We use cooperative game theory to model the P2P energy trading so that all prosumers take part rationally. The simulation is carried out using time series data of houses with varying load profiles, solar generation, and energy storage capacities. The results show 5.81\% of total cost benefit in a year and 10.67\% improvement in the resilience of the system due to P2P energy trading.

\end{abstract}


\section{Introduction}
\label{section:Introduction}

\begin{figure}[H]
  \centering
  \includegraphics[width=2.5in]{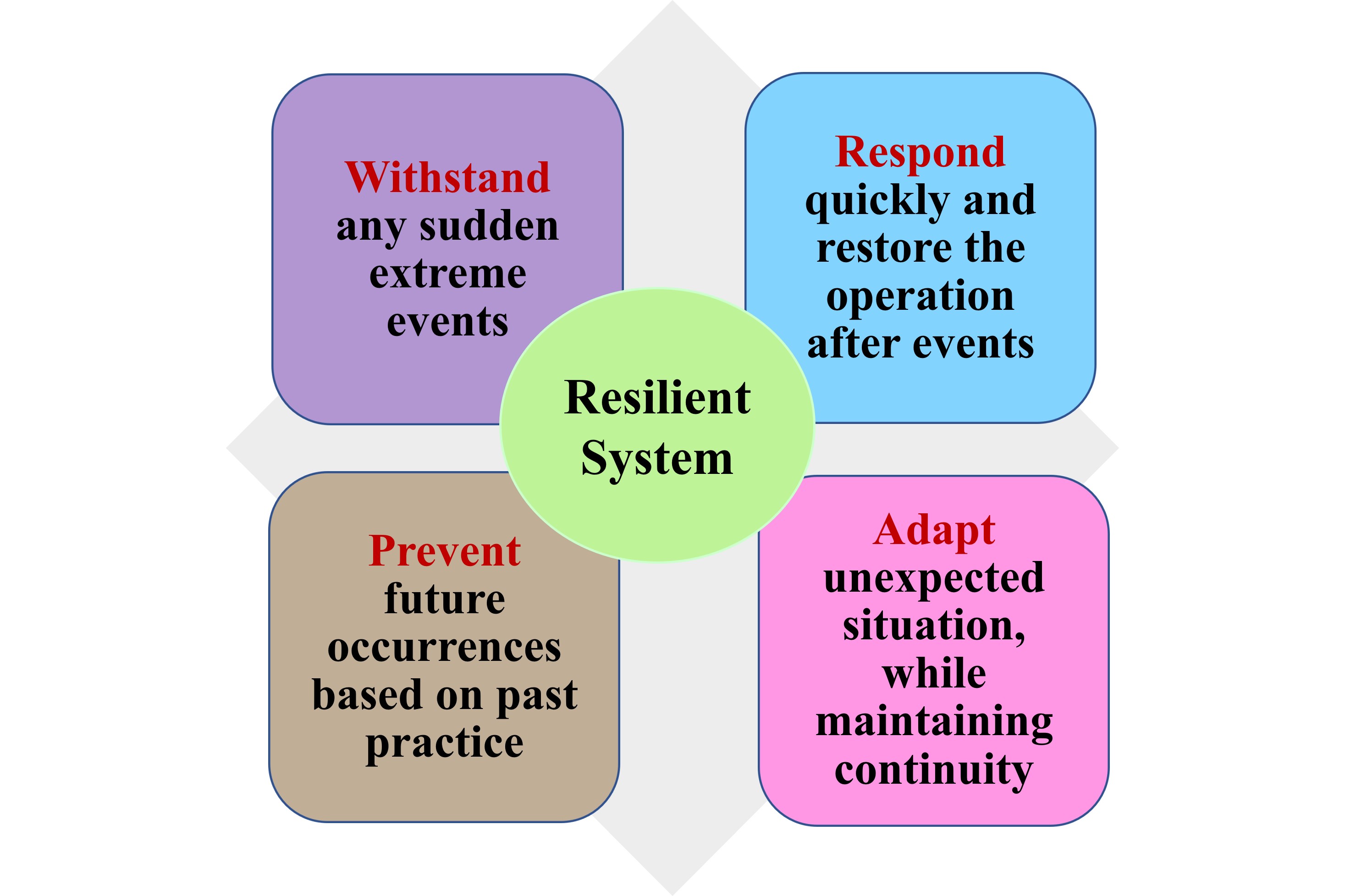}
  \caption{Key aspects of a resilient system.}
  \label{fig:resilient}
\end{figure}

The increasing dependency on the electric power grid, accompanied by more extreme events, emphasizes the need to develop resilient electric power distribution systems \cite{resiliency2}. In 2018, the US federal energy regulatory commission defined resilience as the capability to withstand, respond, adapt and prevent situations such as disruptive events, man-made attacks, and severe technical faults as explained in Fig. \ref{fig:resilient}. \cite{Nationelectricity}. In 2020, Cyclone Nisarga made landfall on the coast of Maharastra, India causing power cuts to more than 2.5 million consumers \cite{Cyclone}. Whenever any such disaster occurs, it affects 90\% of the consumers due to the damages in power distribution feeders \cite{resiliency}. Thus we need to develop resilient systems to handle these events.  

For enhancing the resilience of the system, integrating distributed energy resources (DERs) as local and community resources, microgrid formulation, and line strengthening are the effective solutions \cite{resiliencyoption}. Among these solutions, microgrids for resilience enhancement have gained popularity due to their ability to work in island conditions using tie switches and feed critical loads during extreme events. Thus, microgrids could be used as a local resource or as a community resource to enhance the resilience of the electrical distribution system. Additionally, they could be used as a black-start resource to commence the main generating units during extreme events \cite{microgridoperation}. Microgrid formation in the existing system by various methods is available in the literature, such as re-configurable system design with embedded intelligence and resilient coordination schemes at both local and system levels for tackling extreme events \cite{microgrid1}. There are also some dynamic approaches that assess the resilience and rebuild better resilient grid partitions at run-time \cite{microgrid2},\cite{microgrid3}. Consequently, to enhance resilience, the participation of end users could also account for effectively improving the economic and resilient operation of microgrids; this is not analyzed. For improving the economic situation in a microgrid, peer-to-peer energy trading is best suited. Here consumers who can produce energy will be able to trade energy with their peers and exchange the surplus energy when required. This would result in cost benefit to all consumers taking part in the P2P trading \cite{p2p_resiliency1}, \cite{p2p_resiliency2}.

The idea of sharing resources and services between owners and users was first introduced in 1978 and was termed as sharing economy \cite{Felson}. This has recently attracted interest in the housing and transportation sectors \cite{Percoco,Lutz}. It has also gained the attention of electrical energy users to form peer-to-peer networks with the increase in renewable energy resources \cite{ZHANG}. Game theory provides a mathematical model to analyze the peer-to-peer networks in either cooperative or non-cooperative manner \cite{Chakraborty-S,Kalathil}. In non-cooperative games, the solution is achieved through Nash equilibrium, where each player is assumed to know the equilibrium strategies of the other players, and no player can gain anything by changing their strategy \cite{Book_GT}. In cooperative games, all players cooperate to achieve a common goal by forming coalitions; each player makes contributions that provide strength to the coalition \cite{Poor}. In \cite{Kalathil-S}, solar energy is shared in a non-cooperative manner under a net-metering policy without time-of-use (ToU) pricing. A cooperative strategy for sharing energy storage units under net metering with ToU pricing is discussed in \cite{Victor-GM}. A cooperative model for sharing storage units with cost allocation based on nucleolus that ensures fairness is discussed in \cite{Yang}; only ToU pricing is considered. Sharing of individual energy storage units and a combined storage unit is analyzed for ToU prices using coalitional games in \cite{Chakraborty}. A coalition formation game for P2P energy trading between different types of prosumers is developed in \cite{TUSHAR2}; pricing conditions are not considered. In these works, the cost-benefit analysis is provided, but the energy analysis taking into account the dependency on grid is not considered.

\begin{figure*}[t]
  \centering
  \includegraphics[width=5.5in]{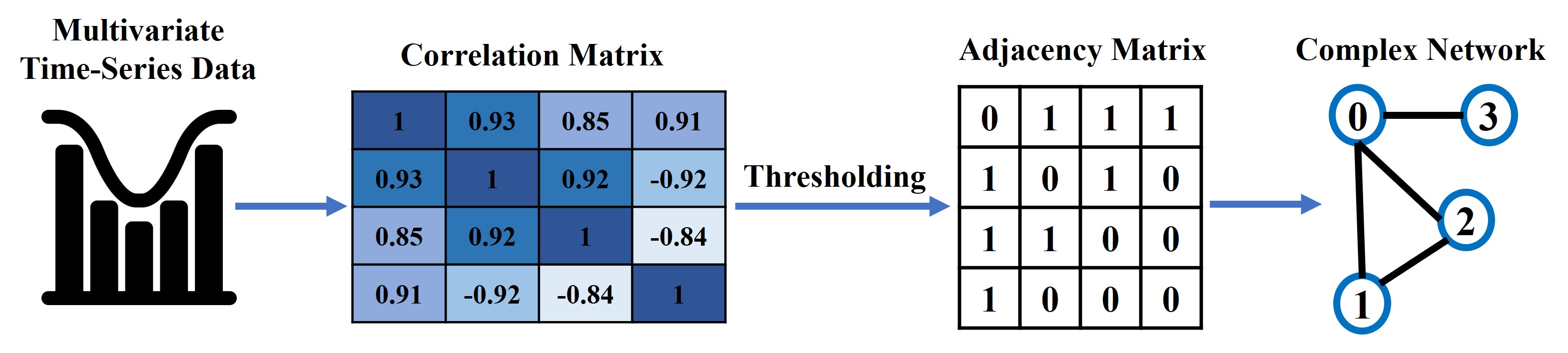}
  \caption{An example for computing complex network from multivariate time-series data using correlation matrix.}
  \label{fig:corr}
\end{figure*}

This paper proposes a novel method through which we can quantify and measure resilience in a microgrid. The analysis would help to understand the impact of DERs and P2P trading in a microgrid. We first consider the standard IEEE 123 node test feeder and divide the loads among residential houses having solar and energy storage units. We then partition the system into 7 microgrids based on the switches. Using a complex network for each microgrid, we compute the percolation threshold and identify the microgrid that is most resilient. We now consider this microgrid to have a P2P network between all the houses. By using cooperative game theory, we build a model for cost-effective sharing. Further, by using a visibility graph for this model, we compute the percolation threshold and compare the resilience with and without sharing of energy resources. Thus, we perform both energy and cost analysis in a P2P network of residential houses and use a quantitative measure to track the resilience of the energy network for all conditions. The main contributions of this work are listed below,

1) We evaluate the energy resilience for all possible partitioned microgrids in the electrical distribution system using complex network by computing the percolation threshold as a quantifiable measure. We consider the most resilient microgrid for further analysis.

2) Using cooperative game theory, we analyze the impact of the P2P energy trading for all houses located in the considered microgrid. The developed model ensures that all houses rationally participate in energy trading to receive economic benefits.

3) We further quantify the improvement of resilience with P2P energy trading for the considered microgrid using visibility graph by computing the percolation threshold.

The rest of the paper is organized as follows; in section \ref{section:Methods} we present how resilience is measured in a distribution system for a single microgrid, and we discuss the cooperative game model for P2P energy trading. In Section \ref{section:Simulation}, we present the simulation study and result analysis with real-world data. Finally, conclusions are drawn in Section \ref{section:Conclusion}.

\section{Materials and Methods}
\label{section:Methods}
In this work, we focus on finding an energy-resilient microgrid in an electrical distribution system and then introduce a peer-to-peer energy trading network within the microgrid. 
\subsection{Resilience in Electrical Distribution System}
\label{subsection:Resiliency}
We use the percolation threshold as a quantifiable measure to evaluate the resilience of an electrical distribution system; this is a mathematical and statistical tool that suggests the phase transition in any network. \cite{resiliency}. The percolation threshold can be computed by finding the percolation strength of a given network using equation (1) \cite{GNN}:\newline

\noindent Percolation Strength,
\begin {equation}
PS_\infty (p) = \frac{1}{VE} \cdot \sum_{m=1}^{X} S_m(p)
\end {equation}
\noindent where, $V$ is the number of nodes/vertices of the network, $E$ is the number of edges of the network, $S (p)$ is the function for bond occupation probability, $p = e/E$; $e$ is the number of edges removed from the initial structure. $S_m (p)$ is computed up to $X$ times (independent), and its cumulative sum is the percolation strength of the network, where $S_m (p)$ is the smallest connected cluster of the network that persists; beyond this, the network would face the transition. Then, we compute the network’s susceptibility using equation (2) as \cite{GNN}:

\begin {equation}
 	\chi(p)=  \frac{(1/V^2 E)\sum_{m=1}^{X} (S_m(p))^2-[PS_\infty (p)]^2} {PS_\infty (p)} 	 
\end {equation}

Finally, the prominent value of $p$ is considered as the percolation threshold $\rho_c$ when the susceptibility reaches its maximum. \newline

\noindent Percolation Threshold, 
\begin {equation}
    \rho_c=arg{[\max \chi(p)]}
\end {equation}

For any electrical system, a higher value of percolation threshold  $\rho_c$ is desired; a higher value signifies higher resilience. It tells that in an electrical network, if a failure occurs in nodes due to natural disasters, man-made attacks, and severe technical faults, then to what extent can the system cope, absorb, and adapt to the changes without interrupting the demand-supply operation?

The first requirement is to create a graph network to compute the percolation threshold. In this work, two graphs are formulated from the time-series data: complex network from correlation matrix and visibility graph.

\subsubsection{Complex network from correlation matrix}
The complex network from correlation matrix gives a measure that correlates various parameters to each other, identifies the strength of the relationship between the parameters, and helps in understanding the complex structures \cite{HAN2021107377}. Complex Network based on correlation matrix for the time-series data helps in suppressing the noises present in the data and provides a robust and reliable network. For two sets of time-series data with $v$ number of nodes, correlation coefficients are computed, and a coefficient matrix of size $v\times v$ is formed to show the relational information among the nodes \cite{percolation1}, \cite{percolation2}. The computed values of these coefficients range between -1 and 1. The computational formula for the Pearson correlation coefficient between the time-series data of nodes $x$ and $y$ is given by:
\begin{equation}
PC= \frac{\sum (P_x^m (t)- \bar P_x^m (t)) \cdot (P_y^m (t)- \bar P_y^m (t))} {\sqrt{\sum (P_x^m (t)
- \bar P_x^m (t))^2 \cdot (P_y^m (t)- \bar P_y^m (t))^2}}
\end{equation}
where $\bar P_x^m$ and $\bar P_y^m$ are the mean for the system parameter of $P_x^m$ and $P_y^m$  respectively.
\begin{figure}[b]
  \centering
  \includegraphics[width=3.3in]{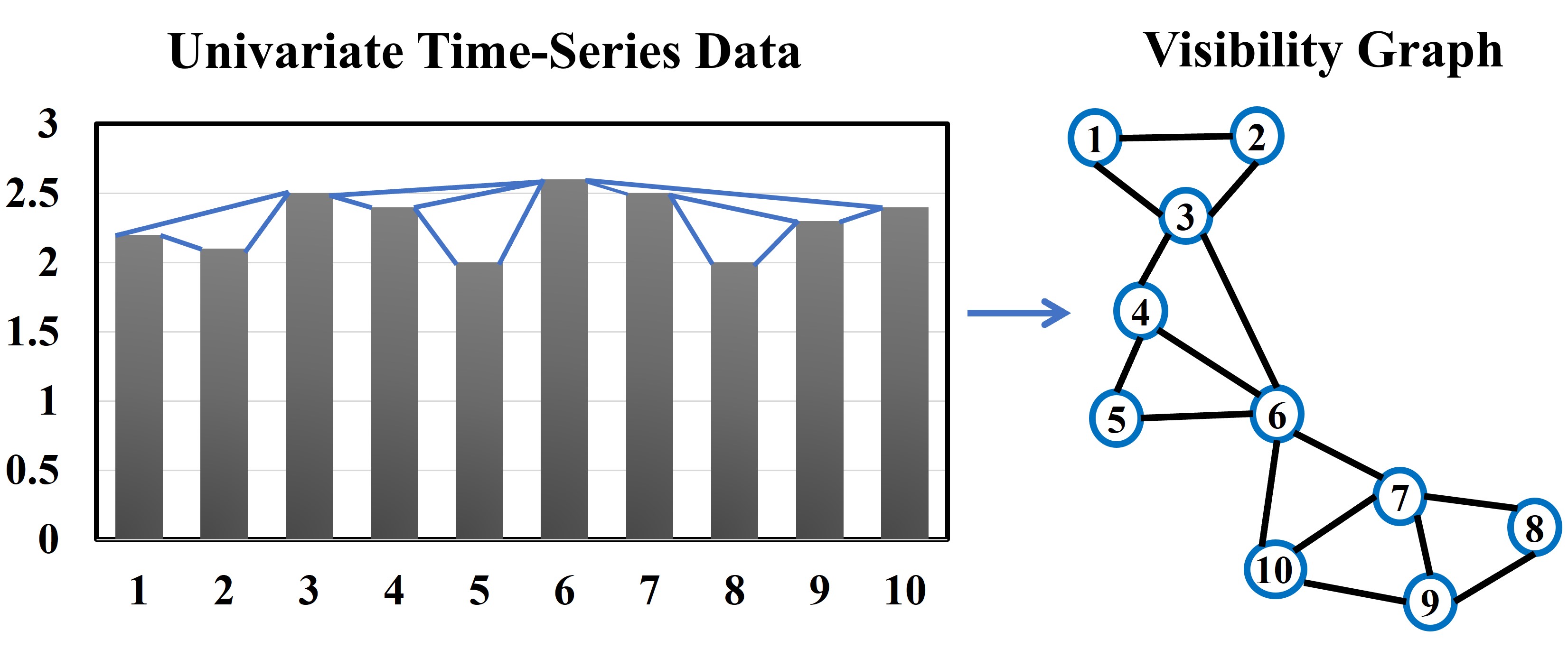}
  \caption{An example for computing complex network from univariate time-series data using visibility graph.}
  \label{fig:visibility}
\end{figure}
After the computation of the Pearson coefficient, a threshold is applied which gives a sparse adjacency matrix to generate the complex network; the process is detailed in Fig. \ref{fig:corr}. 
\subsubsection{Complex network from visibility graph}
Another complex network we created from visibility graph for univariate time series data \cite{Lucas}. These graphs are connected, undirected and invariant under affine transformations of the series data. While creating a visibility graph, every node corresponds in the same order as the input time series data \cite{visibility2}. The connections between two nodes are present if visibility between the corresponding data and the connection line does not intersect any data height, as shown in Fig. \ref{fig:visibility}. Mathematically, this visibility line is drawn between two data points $(x_i,y_i)$ and $(x_j,y_j)$ if other data point $(x_k,y_k)$ satisfies the equation (5).
\begin{equation}
    y_k<y_j+(y_i-y_j) \frac{(x_j-x_k)}{(x_j- x_i )}
\end{equation}
Thus, we created complex networks using two different techniques. For the computation of the percolation threshold, we use equation (1) to compute bond occupation probability for the network by removing the network's edges and determined the percolation strength of the network. Then we compute the network's susceptibility using equation (2). Finally, we identify the bond occupation probability where the susceptibility is maximum and refer to it as the percolation threshold as given in equation (3).

\subsection{Cooperative Game Theory Model for P2P Energy Trading}
\label{subsection:Theoretical Results for the Coalitional Game}

We consider that all houses in the microgrid have made prior investments in both solar PV panels and energy storage units. Each house is indexed by $i \in \mathcal{N} = \{ 1, 2, . . . ,N \} $. We consider a net-metering billing mechanism with two fixed pricing periods for each day: peak ($h$) and off-peak ($l$). The price ($\lambda$) of purchasing from the grid is $\lambda_h$ during the peak period and $\lambda_l$ during the off-peak period. The daily energy consumption of a house during the peak and off-peak periods are $H_{h_i}$ and $H_{l_i}$, respectively. The daily electricity cost of a household without any DERs is
\begin {equation}
C_w(i)=\lambda_{h} H_{h_i} +\lambda_{l} H_{l_i}
\end {equation}
We now consider that each house has invested in energy storage units with capacity $B_i$ and solar PV panels with panel area $a_i$. We consider storage devices and solar PV panels to be ideal. A household's daily solar power generation during the peak and off-peak periods are $G_{h_i}$ and $G_{l_i}$, respectively. Under the net metering billing mechanism, the house is compensated for the net power generation at a price $\mu$ at the end of a billing period. Otherwise, the house would have to pay the net consumption at a price $\lambda$ for the deficit power consumed from the grid. The price ($\mu$) of selling electricity back to the grid for peak and off-peak periods are $\mu_h$ and $\mu_l$, respectively. We consider pricing conditions as $\lambda_h \geq \mu_h$, $\lambda_l \geq \mu_l$, and $\mu_h \geq \lambda_l$. We consider that the house first uses the available energy from both storage and solar units. If there is any deficit $(H_h-aS_h-B)$, it would be purchased from the utility for a peak-period price $\lambda_{h}$. When energy from both the storage and solar units is greater than the consumption of the house, excess energy $(aS_h+B-H_h)$ is sold to the utility for the peak period price $\mu_{h}$. In the off-peak period, we only charge the storage unit to its full capacity and first make use of the available solar energy. If there is any deficit $(H_l+B-aS_l)$, it would be purchased from the utility for a price $\lambda_{l}$. When energy from the solar units is greater than the consumption of the house, excess energy $(aS_l-B-H_l)$ is sold to the utility for a price $\mu_{l}$ Under this scenario \cite{Victor}, the daily cost of the household is
\begin{multline}
    C(i)= \lambda_{h} (H_{h_i}-B_i-G_{h_i})^{+}-\mu_{h}(B_i+G_{h_i}-H_{h_i})^+ \\
    \qquad \qquad \qquad + \lambda_{l}(H_{l_i}+B_i-G_{l_i})^+  -\mu_{l}(G_{l_i}-H_{l_i}-B_i)^+
\end{multline}
where $(x)^+ = \max \{x,0\}$ for any real number $x$.
The houses aggregate their storage and solar PV units and use the aggregated storage capacity to store energy during off-peak periods that they would later use or sell during peak periods. By aggregating their solar and storage units, the unused energy of some houses may be used by others, producing cost savings for the group. We analyze this scenario using cooperative/coalitional game theory \cite{Von}.

\subsubsection{Game Model and Main Results}
We define the coalitional game as $G(\mathcal{N},C)$ with a finite number of prosumers from the set $\mathcal{N}$, each having value function $C$, which is the daily cost of electricity consumption. The prosumers participate in the game to minimize the joint cost and cooperatively share this cost. A coalition is any subset of prosumers $\mathcal{S} \subseteq \mathcal{N}$ where $\mathcal{N}$ is the grand coalition. $H_{h_\mathcal{S}} = \sum_{i\in \mathcal{S}}^{} H_{h_i}$ and $G_{h_\mathcal{S}} = \sum_{i\in \mathcal{S}}^{} G_{h_i}$ denotes the aggregated peak-period house consumption and solar generation. $H_{l_\mathcal{S}} = \sum_{i\in \mathcal{S}}^{} H_{l_i}$ and $G_{l_\mathcal{S}} = \sum_{i\in \mathcal{S}}^{} G_{l_i}$ is the joint off-peak period house consumption and solar generation. The joint storage capacity is $B_\mathcal{S}=\sum_{i\in \mathcal{S}}^{} B_i$. The daily cost of a coalition $\mathcal{S}$ is given by
\begin{multline}
    C(\mathcal{S})= \lambda_{h}(H_{h_\mathcal{S}}-B_\mathcal{S}-G_{h_\mathcal{S}})^+-\mu_{h}(B_\mathcal{S}+G_\mathcal{S}-H_{h_\mathcal{S}})^+\\ \qquad \qquad \quad +\lambda_{l}(H_{l_\mathcal{S}}+B_\mathcal{S}-G_{l_\mathcal{S}})^+ -\mu_{l}(G_\mathcal{S}-H_{h_\mathcal{S}}-B_\mathcal{S})^+
\end{multline}
For a pair of coalitions $\mathcal{S},\mathcal{T} \subset \mathcal{N}$ which are disjoint, i.e., $\mathcal{S}\cap \mathcal{T} = \emptyset $, 
\begin{multline}
    C(\mathcal{T})=\lambda_{h}(H_{h_\mathcal{T}}-B_\mathcal{T}-G_{h_\mathcal{T}})^+
     -\mu_{h}(B_\mathcal{T}+G_{l_\mathcal{T}}-H_{l_\mathcal{T}})^+ \\ 
     \qquad \qquad \quad +\lambda_{l}(H_{l_\mathcal{T}}+B_\mathcal{T}-G_{l_\mathcal{T}})^+
      -\mu_{l}(G_{l\mathcal{T}}-H_{l_\mathcal{T}}-B_\mathcal{T})^+
\end{multline}
\begin{multline}
    C(\mathcal{S}\cup \mathcal{T})=\lambda_{h}(H_{h_\mathcal{S}}H_{h_\mathcal{T}}-B_\mathcal{S}-B_\mathcal{T}-G_{h_\mathcal{S}}-G_{h_\mathcal{T}})^+\\
     \qquad \qquad \qquad -\mu_{h}(B_\mathcal{S}+B_\mathcal{T}+G_{h_\mathcal{S}}+G_{h_\mathcal{T}}-H_{h_\mathcal{S}}-H_{h_\mathcal{T}})^+\\ 
     \qquad \qquad \qquad +\lambda_{l}(H_{l_\mathcal{S}}H_{l_\mathcal{T}}+B_\mathcal{S}+B_\mathcal{T}-G_{l_\mathcal{S}}-G_{l_\mathcal{T}})^+\\ 
     \qquad \qquad  \qquad -\mu_{l}(G_{l_\mathcal{S}}G_{l_\mathcal{T}}-H_{l_\mathcal{S}}-H_{l_\mathcal{T}}-B_\mathcal{S}-B_\mathcal{T})^+
\end{multline}
It is simple to show that the cooperative game $G(\mathcal{N},J)$ for sharing of energy in a P2P network is subadditive, i.e., it satisfies the condition $C(\mathcal{S})+C(\mathcal{T}) \ge C(\mathcal{S} \cup \mathcal{T})$. Therefore, cooperation is advantageous to the players in the game. But we also need to check if the game is stable. In this game, once the grand coalition is formed, players should not break it and be more profitable by forming a coalition with a subset of players. Mathematically, the condition is called balancedness \cite{Saad}. If $\alpha$ is a positive number, then we could show that
\begin{multline}
C(\alpha {S})= \alpha \bigg[ \lambda_h (H_{h_\mathcal{S}} -B_\mathcal{S} -G_{h_\mathcal{S}})^+\\
  \qquad - \mu_h (B_\mathcal{S}+G_{h_\mathcal{S}}-H_{h_\mathcal{S}})^+\\
  \qquad \qquad \qquad + \lambda_l (H_{l_\mathcal{S}} +B_\mathcal{S}-G_{l_\mathcal{S}})^+\\
   \qquad \qquad \qquad \qquad - \mu_l (G_{l_\mathcal{S}} -H_{l_\mathcal{S}}- B_\mathcal{S})^+\bigg] 
\end{multline}
This shows us that $C(\alpha \mathcal{S}) = \alpha C(\mathcal{S})$, therefore, $C$ is a positive homogeneous function. And if $\alpha $ be any balanced map such that $\alpha : 2^\mathcal{N} \rightarrow [0,1] $, then $\underset{S \in 2^\mathcal{N} }{\sum}\alpha(\mathcal{S}) \mathbf{1}_\mathcal{S}(i) =1$ where $\mathbf{1}_\mathcal{S}$ is an indicator function of set $\mathcal{S}$, i.e., $\mathbf{1}_\mathcal{S}(i)=1$ if $i\in \mathcal{S}$ and $\mathbf{1}_\mathcal{S}(i)=0$ if $i \notin \mathcal{S}$. As the cost $C$ is a homogeneous function and the game is also subadditive, we can write,
\begin{multline}
\sum_{\mathcal{S}\in 2^\mathcal{N}}^{} \alpha(S) C(\mathcal{S}) =\underset{\mathcal{S} \in 2^\mathcal{N}}{\sum} C\bigg(\alpha(\mathcal{S}) H_{h_\mathcal{S}}, \alpha(\mathcal{S}) G_{h_\mathcal{S}}, \\ \qquad \qquad \qquad \qquad \alpha (\mathcal{S}) B_\mathcal{S}, \alpha(\mathcal{S}) H_{l_\mathcal{S}}, \alpha(\mathcal{S}) G_{l_\mathcal{S}}\bigg)\\
= C \bigg( \sum\limits_{\substack{i \in \mathcal{N}}} \sum\limits_{\substack{\mathcal{S} \in 2^\mathcal{N}}}\alpha(\mathcal{S}) \mathbf{1}_\mathcal{S}(i) H_{h_i},\sum\limits_{\substack{i \in \mathcal{\mathcal{N}}}} \sum\limits_{\substack{\mathcal{S} \in 2^N}} \alpha(\mathcal{S}) \mathbf{1}_\mathcal{S}(i) G_{h_i},\\ 
\qquad \sum\limits_{\substack{i \in \mathcal{N}}} \sum\limits_{\substack{\mathcal{S} \in 2^\mathcal{N}}} \alpha(\mathcal{S}) \mathbf{1}_\mathcal{S}(i) B_{i},\sum\limits_{\substack{i \in \mathcal{N}}} \sum\limits_{\substack{\mathcal{S} \in 2^\mathcal{N}}}\alpha(\mathcal{S}) \mathbf{1}_\mathcal{S}(i) H_{l_i},\\
\sum\limits_{\substack{i \in \mathcal{\mathcal{N}}}} \sum\limits_{\substack{\mathcal{S} \in 2^N}} \alpha(\mathcal{S}) \mathbf{1}_\mathcal{S}(i) G_{l_i} \bigg)\\
=C(H_{h_\mathcal{N}},G_{h_\mathcal{N}},B_\mathcal{N},H_{l_\mathcal{N}},G_{l_\mathcal{N}})=C(\mathcal{N}) \qquad \qquad \qquad
\end{multline}
where $C(\mathcal{N})$ is the cost of the grand coalition defined as
\begin{multline}
    C(\mathcal{N})= \lambda_{h}(H_{h_\mathcal{N}}-B_\mathcal{N}-G_{h_\mathcal{N}})^+
    -\mu_{h}(B_\mathcal{N}+G_\mathcal{N}-H_{h_\mathcal{N}})^+\\ 
    \qquad \qquad +\lambda_{l}(H_{l_\mathcal{N}}+B_\mathcal{N}-G_{l_\mathcal{N}})^+ -\mu_{l}(G_{l\mathcal{N}}-H_{l_\mathcal{N}}-B_\mathcal{N})^+
\end{multline}
This shows that the game G$(\mathcal{N},C)$ is balanced. Thus the game is profitable and stable. A grand coalition will be formed, and prosumers will not break the coalition rationally. Now, the joint cost of the grand coalition needs to be allocated to the individual agents. Let $\xi(i)$ denote the cost allocation for prosumer $i \in \mathcal{S} $. For a coalition, $\mathcal{S}$, $\xi(\mathcal{S}) = \underset{i\in \mathcal{S}}{\sum} \xi(i)$ is the sum of cost allocations of all members of the coalition. The cost allocation is said to be an imputation if it is simultaneously efficient $(C(\mathcal{S}) =\xi(\mathcal{S}))$ and individually rational $(C(i) \ge \xi(i) )$ \cite{CHURKIN}. Let $\mathcal{I}$ denote the set of all imputations. The core, $\mathcal{C}$ of the coalition game $G(\mathcal{N},C)$ \cite{CHURKIN} includes all cost allocations from set $\mathcal{I}$ such that cost of no coalition is less than the sum of allocated costs of all prosumers, it is defined as follows:
\begin{equation}
\mathcal{C} = \bigg(\xi \in \mathcal{I} : C(\mathcal{S}) \ge \xi(\mathcal{S}), \forall \mathcal{S} \in 2^\mathcal{N} \bigg)
\end{equation}
According to Bordareva-Shapley value theorem \cite{Saad}, the coalitional game has a non-empty core if it is balanced. Since our game is balanced, the core is non-empty; hence, it is possible to find a cost allocation in the core of the coalition game. We develop a cost allocation {$\xi_i$} with an analytical formula that is straightforward to compute.
\begin{equation}
\resizebox{1\hsize}{!}{$
\xi(i)=\left\{
    \begin{aligned}
       &K_i \quad \textrm{if} \quad  H_{h_\mathcal{N}} \ge B_\mathcal{N}+G_{h_\mathcal{N}} \; \& \; H_{l_\mathcal{N}}+B_\mathcal{N} \ge G_{l_\mathcal{N}}\\ 
        &L_i \quad \textrm{if} \quad  H_{h_\mathcal{N}} < B_\mathcal{N}+G_{h_\mathcal{N}} \; \& \; H_{l_\mathcal{N}}+B_\mathcal{N} \ge G_{l_\mathcal{N}}\\ 
        &M_i \quad \textrm{if} \quad  H_{h_\mathcal{N}} \ge B_\mathcal{N} +G_{h_\mathcal{N}} \; \& \; H_{l_\mathcal{N}}+B_\mathcal{N} < G_{l_\mathcal{N}}\\ 
        &N_i \quad \textrm{if} \quad  H_{h_\mathcal{N}} < B_\mathcal{N}+G_{h_\mathcal{N}} \; \& \; H_{l_\mathcal{N}}+B_\mathcal{N} < G_{l_\mathcal{N}}&& 
    \end{aligned}\right.$}
\end{equation}
\noindent where,
\begin{flalign*}
 K_i&=\lambda_{b_i}B_i+ \lambda_{a_i}a_i+ \lambda_h (H_{h_i}-B_i-G_{h_i})\\
        & \qquad \qquad \qquad \qquad \qquad \qquad+\lambda_l (H_{l_i}+B_i-G_{l_i})\\
L_i&=\lambda_{b_i}B_i+ \lambda_{a_i}a_i -\mu_h (B_i+G_{h_i}-H_{h_i})\\
        & \qquad \qquad \qquad \qquad \qquad \qquad + \lambda_{l}(H_{l_i}+B_i-G_{l_i})\\
 M_i&=\lambda_{b_i}B_i+ \lambda_{a_i}a_i+ \lambda_{h} (H_{h_i}-B_i-G_{h_i})\\
        & \qquad \qquad \qquad \qquad \qquad \qquad -\mu_l(G_{l_i}-H_{l_i}-B_i)\\
 N_i&=\lambda_{b_i}B_i+ \lambda_{a_i}a_i -\mu_h (B_i+G_{h_i}-H_{h_i})\\ 
        & \qquad \qquad \qquad \qquad \qquad \qquad -\mu_l(G_{l_i}-H_{l_i}-B_i)&& 
\end{flalign*} 
\noindent The cost of the grand coalition is 
\begin{equation}
\resizebox{1\hsize}{!}{$
C(\mathcal{N})=\left\{ 
    \begin{aligned}
       &K_\mathcal{N} \quad \textrm{if} \quad  H_{h_\mathcal{N}} \ge B_\mathcal{N}+G_{h_\mathcal{N}} \; \& \; H_{l_\mathcal{N}}+B_\mathcal{N} \ge G_{l_\mathcal{N}}\\ 
        &L_\mathcal{N} \quad \textrm{if} \quad  H_{h_\mathcal{N}} < B_\mathcal{N}+G_{h_\mathcal{N}} \; \& \; H_{l_\mathcal{N}}+B_\mathcal{N} \ge G_{l_\mathcal{N}}\\ 
        &M_\mathcal{N} \quad \textrm{if} \quad  H_{h_\mathcal{N}} \ge B_\mathcal{N} +G_{h_\mathcal{N}} \; \& \; H_{l_\mathcal{N}}+B_\mathcal{N} < G_{l_\mathcal{N}}\\ 
        &N_\mathcal{N} \quad \textrm{if} \quad  H_{h_\mathcal{N}} < B_\mathcal{N}+G_{h_\mathcal{N}} \; \& \; H_{l_\mathcal{N}}+B_\mathcal{N} < G_{l_\mathcal{N}}&& 
    \end{aligned}\right.$}
\end{equation}
\noindent where,
\begin{flalign*}
 K_\mathcal{N}&=\underset{i\in \mathcal{N}}{\sum}(\lambda_{b_i}B_i+ \lambda_{a_i}a_i )
         +\lambda_h (H_{h_\mathcal{N}}-B_\mathcal{N}-G_{h_\mathcal{N}}) \\
        & \qquad \qquad \quad  \qquad \qquad+\lambda_l (H_{l_\mathcal{N}}+B_\mathcal{N}-G_{l_\mathcal{N}})\\
L_\mathcal{N}&=\underset{i\in \mathcal{N}}{\sum}(\lambda_{b_i}B_i+ \lambda_{a_i}a_i )
         -\mu_h (B_\mathcal{N}+G_{h_\mathcal{N}}-H_{h_\mathcal{N}})\\
        & \qquad \qquad \quad \qquad \qquad \qquad + \lambda_{l}(H_{l_\mathcal{N}}+B_\mathcal{N}-G_{l_\mathcal{N}}) \\
 M_\mathcal{N}&=\underset{i\in \mathcal{N}}{\sum}(\lambda_{b_i}B_i +\lambda_{a_i}a_i)
         +\lambda_{h} (H_{h_\mathcal{N}}-B_\mathcal{N}-G_{h_\mathcal{N}})\\
        & \qquad \qquad \quad \qquad \qquad \qquad -\mu_l(G_{l_\mathcal{N}}-H_{l_\mathcal{N}}-B_\mathcal{N}) \\
 N_\mathcal{N}&=\underset{i\in \mathcal{N}}{\sum}(\lambda_{b_i}B_i+ \lambda_{a_i}a_i)
         -\mu_h (B_\mathcal{N}+G_{h_\mathcal{N}}-H_{h_\mathcal{N}})\\ 
        & \qquad \qquad \quad \qquad \qquad \qquad -\mu_l(G_{l_\mathcal{N}}-H_{l_\mathcal{N}}-B_{\mathcal{N}})&& 
\end{flalign*} 
The cost of an individual household without joining the coalition is 
\begin{equation}
\resizebox{1\hsize}{!}{$
C(i)=\left\{ 
    \begin{aligned}
       &K_i \quad \textrm{if} \quad  H_{h_i} \ge B_i+G_{h_i} \; \& \; H_{l_i}+B_i \ge G_{l_i}\\ 
        &L_i \quad \textrm{if} \quad  H_{h_i} < B_i+G_{h_i} \; \& \; H_{l_i}+B_i \ge G_{l_i}\\ 
        &M_i \quad \textrm{if} \quad  H_{h_i} \ge B_i +G_{h_i} \; \& \; H_{l_i}+B_i < G_{l_i}\\ 
        &N_i \quad \textrm{if} \quad  H_{h_i} < B_i+G_{h_i} \; \& \; H_{l_i}+B_i < G_{l_i}&& 
    \end{aligned}\right.$}
\end{equation}
For all four cases: $H_{h_\mathcal{N}} \ge B_\mathcal{N}+G_{h_\mathcal{N}} \; \& \; H_{l_\mathcal{N}}+B_\mathcal{N} \ge G_{l_\mathcal{N}}$, $H_{h_\mathcal{N}} < B_\mathcal{N}+G_{h_\mathcal{N}} \; \& \; H_{l_\mathcal{N}}+B_\mathcal{N} \ge G_{l_\mathcal{N}}$, $H_{h_\mathcal{N}} \ge B_\mathcal{N} +G_{h_\mathcal{N}} \; \& \; H_{l_\mathcal{N}}+B_\mathcal{N} < G_{l_\mathcal{N}}$, and $ H_{h_\mathcal{N}} < B_\mathcal{N}+G_{h_\mathcal{N}} \; \& \; H_{l_\mathcal{N}}+B_\mathcal{N} < G_{l_\mathcal{N}}$, the cost allocation $ \xi(i) :i \in \mathcal{N}$ satisfies the budget balance, i.e., $\underset{i\in \mathcal{N}}{\sum} \xi(i) = C(\mathcal{N}) $. And the cost allocation is individually rational i.e., $\xi(i) \le C(i)$  for all $i \in N$ and for all the given conditions: $H_{h_i} \ge B_i+G_{h_i} \; \& \; H_{l_i}+B_i \ge G_{l_i}, H_{h_i} < B_i+G_{h_i} \; \& \; H_{l_i}+B_i \ge G_{l_i}, H_{h_i} \ge B_i +G_{h_i} \; \& \; H_{l_i}+B_i < G_{l_i}, H_{h_i} < B_i+G_{h_i} \; \& \; H_{l_i}+B_i < G_{l_i}$. Thus the cost allocation is an imputation, and we can say that the imputation $\xi(i)$ belongs to the core of the cooperative game as $\underset{i\in \mathcal{S}}{\sum} \xi(i) \le C(\mathcal{S})$ for the coalition $\mathcal{S}\subseteq \mathcal{N}$.

\subsubsection{Sharing Mechanism}
For the cost allocation $\xi(i)$ to belong to the core, we consider the price of selling or buying energy in a P2P network by all the prosumers to be $\pi_h$ for the peak period and $\pi_l$ for the off-peak period. There are two conditions for which the price of energy sharing varies; this depends on the total energy of the network and the individual energy of a house. 

The first condition is when the total energy consumption of the P2P network is greater than the total energy generation. In this condition, if all houses have their individual consumption more than their generation, then there would be no sharing of energy in the P2P network. If there are few houses that have excess energy after catering to their own consumption, then they can sell their excess energy first to the other houses and the remainder to the grid. The price of buying/selling energy in the P2P network under this condition is considered to be $\pi_h=\lambda_h$ for the peak period and $\pi_l=\lambda_l$ for the off-peak period.

The second condition is when the total energy consumption of the P2P network is lower than the total energy generation. In this condition, if all houses have their individual consumption lower than their generation, then there would be no sharing of energy in the P2P network. If there are few houses that have deficit energy after utilizing their renewable energy, then they can buy the required energy from the other houses. The houses supplying this deficit will first sell to their peers and sell the remainder to the grid. The price of buying/selling energy in the P2P network under this condition is considered as $\pi_h=\mu_h$ for the peak period and $\pi_l=\mu_l$ for the off-peak period.

\begin{figure}[b]
  \centering
  \includegraphics[width=3.4in]{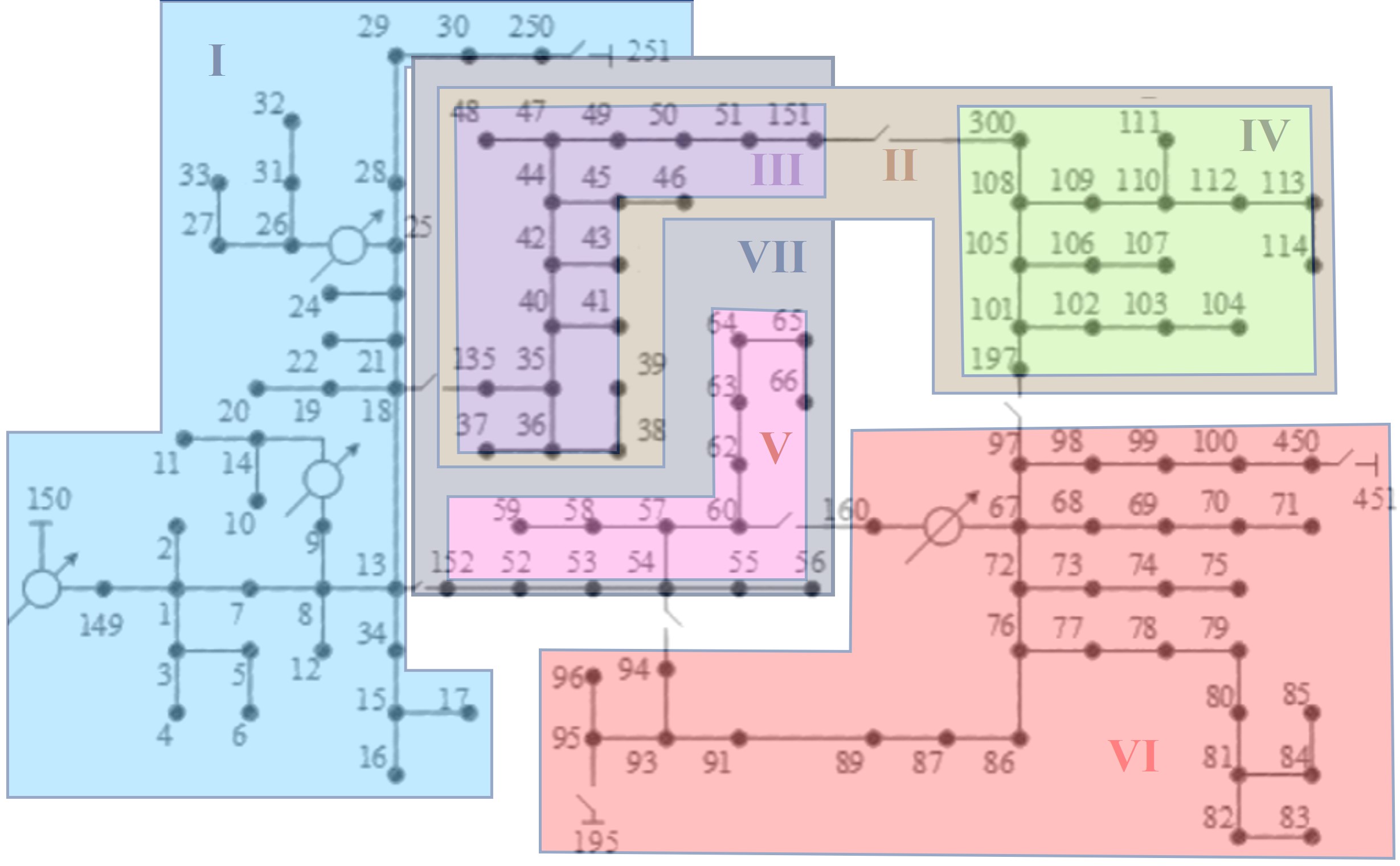}
  \caption{IEEE-123 bus test feeder partitioned into seven microgrids considering switching topology.}
  \label{fig:IEEE-123}
\end{figure}

\begin{figure}[]
  \centering
  \includegraphics[width=3in]{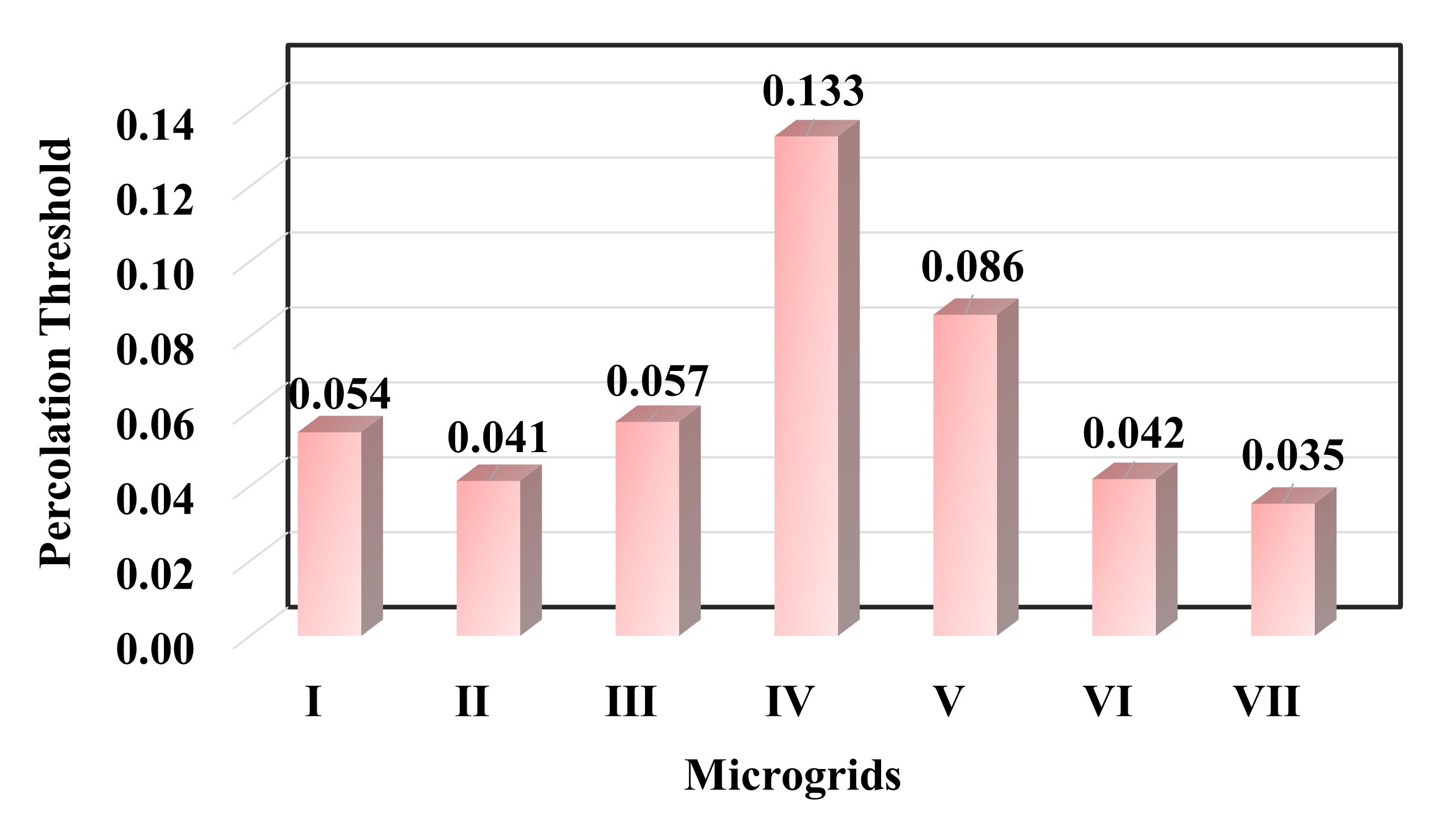}
  \caption{Percolation Threshold of seven microgrids computed from complex network.}
  \label{fig:PT_7_MGs}
\end{figure}

\begin{figure}[b]
  \centering
  \includegraphics[width=2.75in]{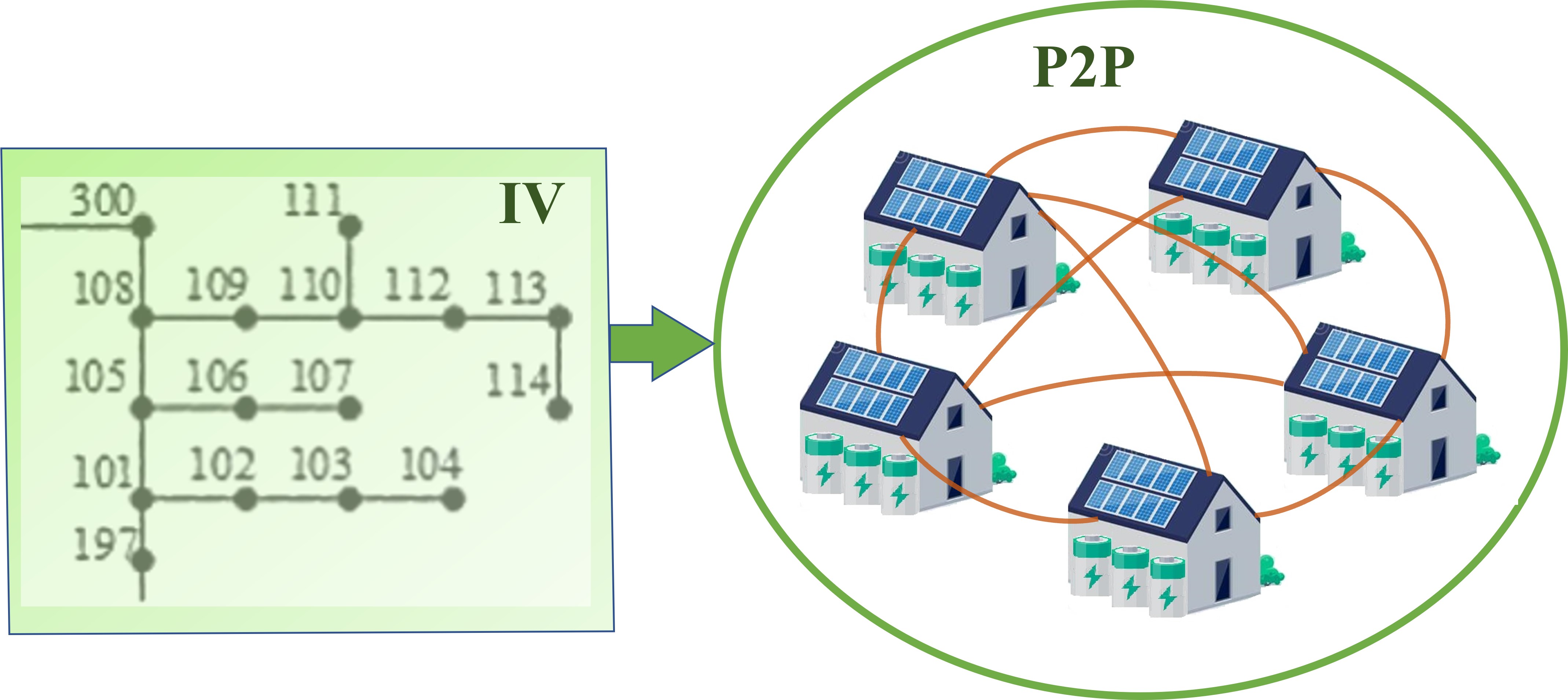}
  \caption{P2P network in Microgrid-IV comprising 48 houses.}
  \label{fig:MG4_P2P}
\end{figure}

\section{Simulation Study, Result Analysis and Discussion}
\label{section:Simulation}

\subsection{Data Processing}
The simulation model is developed in GridLAB-D, an open-source software that provides a platform for distribution system design with the incorporation of renewable energy sources \cite{resiliency}. There are 85 constant loads with a total real-power requirement of 3620.5 kW in the IEEE-123 node test feeder system. We introduced 518 houses of different floor areas with varying load profiles, solar PV panels with a size of approximately 10\% of floor area \cite{Victor}, and random storage capacity. We simulated the model for 365 days with four-hour intervals considering the climate profile of Bakersfield, California. We obtained the time-series data of real power, energy consumption, and generation for two conditions, i.e., with and without renewable energy sources. This data is used for further analysis.


\begin{figure*}[]
  \centering
  \includegraphics{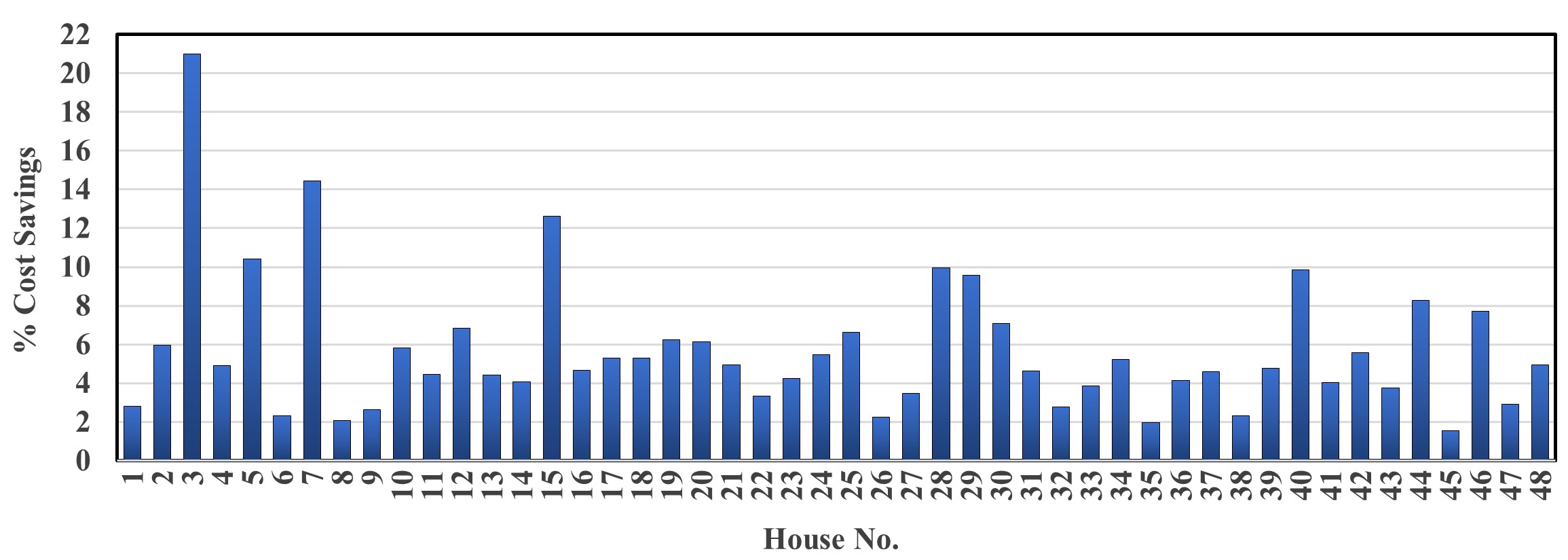}
  \caption{Cost savings of all houses for one year.}
  \label{fig:Houses_save_comp}
\end{figure*}

\begin{figure*}[]
  \centering
  \includegraphics{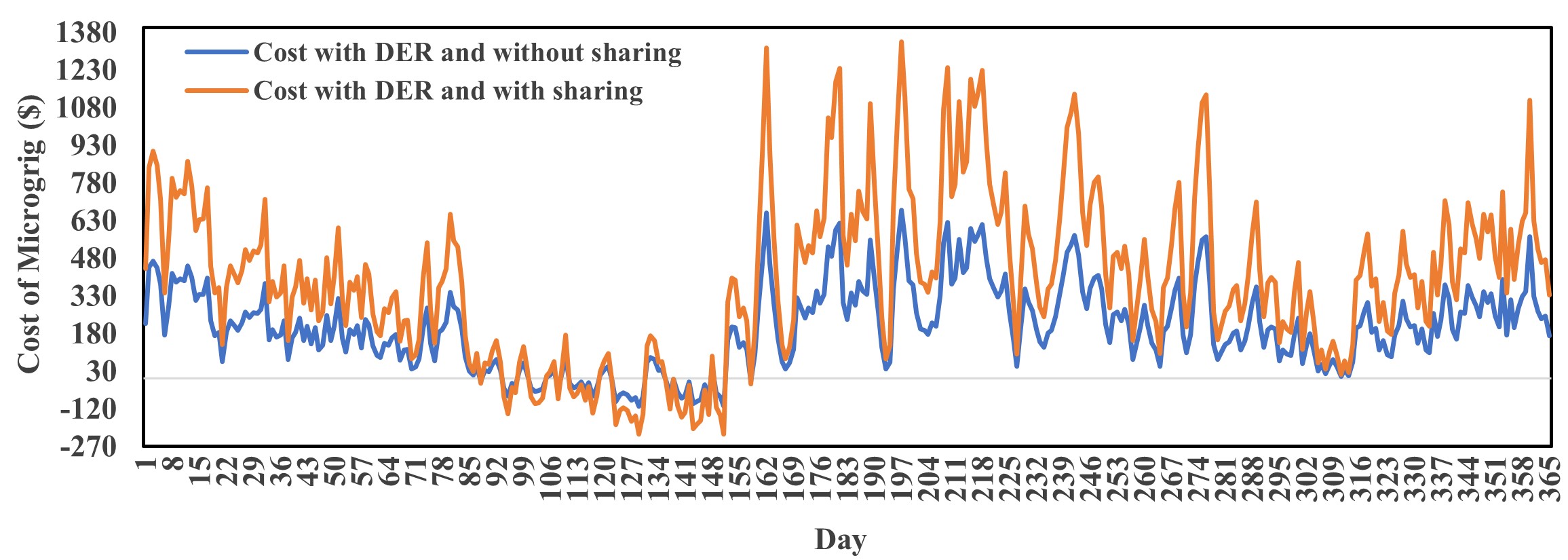}
  \caption{Total cost of entire microgrid for one year.}
  \label{fig:MG_cost_comp}
\end{figure*}

For the P2P trading scenario, we have considered a net-metering billing mechanism with time-of-use pricing scheme (ToU); peak period price and off-peak period price for buying (54\mbox{\textcentoldstyle}/kWh and 22\mbox{\textcentoldstyle}/kWh) and selling (30\mbox{\textcentoldstyle}/kWh and 13\mbox{\textcentoldstyle}/kWh) energy to and from the grid. Prices for trading only within the P2P network are given by the conditions for $\pi_h$ and $\pi_l$.  We have considered peak period from 8hrs to 20hrs and off-peak period from 20hrs to 8hrs. In the peak period, the entire storage energy is utilized. And in the off-peak period, the storage is charged to its total capacity for all houses. The solar energy is first utilized for the respective house, and then excess energy is either sold to the grid or P2P network for both periods. Using cooperative game theory, we have shown that all houses would rationally choose to connect to the P2P network and form a stable coalition with good cooperative behavior providing economic benefits for all houses. 

\subsection{Result Analysis and Discussion}
\label{section:Results}

Considering the switching operation of 11 switches present in the IEEE 123 test network to form the various combinations to partition system into microgrids. We identified 7 self-sufficient and reliable microgrids with the integration of renewable energy sources as shown in Fig. \ref{fig:IEEE-123}. For the evaluation of resilience, we computed the complex network from correlation matrix for each microgrid combination. These complex networks are created for each microgrid with the time-series data of real power for the conditions, with and without DERs. And then the percolation threshold is computed as shown in Fig. \ref{fig:PT_7_MGs}. The computation of percolation threshold values is done by taking cross-correlation among data into consideration. The higher the percolation threshold, the more would be the resilience of the microgrid. From Fig. \ref{fig:PT_7_MGs} we could say that Microgrid-IV is more resilient in comparison to other microgrids as it has a percolation threshold value of 0.133; we further considered this microgrid for P2P trading, as shown in Fig. \ref{fig:MG4_P2P}. This microgrid comprises a total of 48 houses that would participate in P2P energy trading.

In Fig. \ref{fig:Houses_save_comp} we have shown the computed percentage savings of each house for two cases, with and without sharing of DERs. The descriptive statistics of the total yearly cost and savings of all 48 houses in a year are given in Table \ref{tab:Des_cost}. The maximum savings is 20.98\% for House 3 and the minimum is 1.53\% for House 45, the average savings per house is 5.67\%. In Fig. \ref{fig:MG_cost_comp}, we observe how the cost of the entire microgrid of 48 houses varies daily for a period of one year, and we have tabulated the monthly cost in Table \ref{tab:Mon_cost}. The cost savings is very low for the summer months of April and May due to more solar generation and lesser energy requirement of each individual house, but for all the remaining months the monthly average in cost savings is 5.5\% with January having the highest of 7.99\% and October have the lowest of 3.5\%.

\begin{table}[]
\centering
\setlength{\tabcolsep}{1pt}
\caption{Descriptive Statistics of Total Cost and Savings for all 48 Houses in a Year}
\begin{tabular}{>{\centering}m{5em} c c c c c}
    \toprule
    \multicolumn{1}{r}{} & \multicolumn{1}{>{\centering}m{4.5em}}{Without DER(\$)} & \multicolumn{1}{>{\centering}m{4.5em}}{Without sharing(\$)} & \multicolumn{1}{>{\centering}m{4.5em}}{With sharing(\$)} & \multicolumn{1}{>{\centering}m{4.5em}}{Savings(\$)} & \multicolumn{1}{>{\centering}m{4.5em}}{\%Savings}\\
\midrule
Mean      & 5427     & 1957      & 1857     & 100  & 5.67  \\
Minimum     & 2949     & 274      & 216      & 28   & 1.53   \\
1\textsuperscript{st} quartile & 4700    & 1222    & 1166     & 48   & 3.68 \\
Median   & 5239   & 1638     & 1584   & 87   & 4.84 \\
3\textsuperscript{rd} quartile & 6146    & 2524    & 2413     & 131  & 6.34  \\
Maximum    & 8374  & 4570   & 4346  & 229  & 20.98   \\
\bottomrule
\end{tabular}
\label{tab:Des_cost}
\end{table}

 \begin{table}[]
\caption{Cost Comparison of Microgrid}
\setlength{\tabcolsep}{1pt}
\begin{tabular}{>{\centering}m{5em} c c c c c c}
    \toprule
    \multicolumn{1}{>{\centering}m{4.5em}}{Month} & \multicolumn{1}{>{\centering}m{4.5em}}{Without DER(\$)} &
    \multicolumn{1}{>{\centering}m{4.5em}}{Without sharing(\$)} & \multicolumn{1}{>{\centering}m{4.5em}}{With sharing(\$)} & \multicolumn{1}{>{\centering}m{4.5em}}{Savings(\$)} & \multicolumn{1}{>{\centering}m{4.5em}}{\%Savings}\\
\midrule
Jan   & 17984  & 9263  & 8522  & 740  & 7.99  \\
Feb   & 13605  & 5276  & 5018  & 258  & 4.90  \\
Mar   & 13562  & 4035  & 3871  & 164  & 4.08  \\
Apr   & 10223  & -115  & -116  & 1.01    & 0.87 \\
May   & 10719  & -1111 & -1114 & 2.37    & 0.21 \\
Jun   & 24194  & 8253  & 7815  & 438  & 5.32  \\
Jul   & 27504  & 10665 & 10071 & 594  & 5.57  \\
Aug   & 27783  & 11089 & 10574 & 515  & 4.65  \\
Sep   & 21762  & 7245  & 6728  & 517  & 7.14  \\
Oct   & 18785  & 6317  & 6096  & 220  & 3.50  \\
Nov   & 13076  & 4139  & 3962  & 177  & 4.29  \\
Dec   & 17534  & 8776  & 8116  & 660  & 7.52  \\
\textbf{Total} & \textbf{216736} & \textbf{73836} & \textbf{69545} & \textbf{4291} & \textbf{5.81} \\
\bottomrule
\end{tabular}
\label{tab:Mon_cost}
\end{table}

\begin{table}[]
\centering
\caption{Percolation Threshold values computed from Energy taken from Grid.}
\begin{tabular}{>{\centering}m{6em} c c c}
    \toprule
    \multicolumn{1}{>{\centering}m{6em}}{Cases}  &
    \multicolumn{1}{>{\centering}m{6em}}{Energy taken from grid (kWh)} & \multicolumn{1}{>{\centering}m{6em}}{Percolation Threshold}\\
\midrule
Microgrid without P2P  & 492,483 & 0.22653  \\
Microgrid with P2P  & 474,631 & 0.25077  \\
\bottomrule
\end{tabular}
\label{tab:percolation_visibility}
\end{table}

From the analysis, we observed that introducing P2P energy trading in a resilient microgrid result in cost-effective operation. We have also analyzed the energy in the microgrid for a year and computed the energy taken from the grid for two cases, i.e., when the microgrid does and does not engage in P2P trading. Then for both cases, time-series data is considered for creating a visibility graph from which the percolation threshold is computed. The computed percolation threshold is based on auto-correlation between the univariate data. The resilience of the microgrid is computed to show the impact of P2P trading as tabulated in Table \ref{tab:percolation_visibility}. We can observe that the percolation threshold value is increased from 0.226 to 0.250 when P2P energy trading takes place, and the energy taken from the grid is reduced by 17,852 kWh. Thus, peer-to-peer energy trading improves the economic and energy resilience condition of the microgrid.

\section{Conclusion}
\label{section:Conclusion}

A peer-to-peer energy trading model for a microgrid is developed and analyzed for energy resilience and economic benefits. We simulated the standard IEEE 123 test bus feeder integrated with solar PV panels and battery storage. We identified the possible combinations for forming microgrids by considering the switching operation. The complex network from correlation matrix is created for microgrids to identify the network strength, and we computed the percolation threshold to determine the coping capability of networks during extreme events. The more the percolation threshold’s value more is the resilience of the network. Further, we introduced energy sharing between residential houses in a community under the net metering billing mechanism with a time-of-use pricing scheme. We then used a cooperative game theory approach to model energy sharing in a peer-to-peer network for profitability and stability. We developed a sharing mechanism and a cost allocation rule such that all houses profit through either buying from or selling to the P2P network. Thus, our results show that sharing energy in a resilient microgrid provides cost savings for all the houses. On the community level, the total cost savings is 5.81\% for a year. On an individual household level, 27 houses have savings of less than 5\% for a year, 17 houses save between 5\% to 10\%, and 4 houses save more than 10\%.

Further, we computed a percolation threshold value for the total energy taken from the grid when the microgrid does and does not participate in P2P trading. It was observed that the percolation threshold value increased from 0.226 to 0.250, which accounted for a 10.67\% improvement in resilience when the microgrid engaged in P2P trading. Thus, we identified a resilient microgrid in the electrical distribution system and implemented P2P energy trading, resulting in cost savings and further enhancing the microgrid’s resilience. In future work, we will be extending our work towards analyzing the behavior of other microgrids in the network when P2P trading is introduced and comparing the resilience and cost-effectiveness.

\bibliographystyle{IEEEtran}
\bibliography{main}

\end{document}